\documentclass[a4paper]{article}
 \pdfoutput=1
\usepackage{todonotes}
\usepackage{multirow}
\usepackage{INTERSPEECH2021}
\usepackage{nidanfloat}
\usepackage{enumitem}
\usepackage{float}
\usepackage{graphicx}
 \usepackage{caption}

\title{Dissecting User-Perceived Latency of On-Device E2E Speech Recognition}
\name{Yuan Shangguan, Rohit Prabhavalkar, Hang Su, Jay Mahadeokar, Yangyang Shi, Jiatong Zhou, Chunyang Wu, Duc Le, Ozlem Kalinli, Christian Fuegen, Michael L. Seltzer}

\address{Facebook AI, USA}
\email{\{yuansg,prabhavalkar\}@fb.com}

\begin{document}

\maketitle
\begin{abstract}
  
As speech-enabled devices such as smartphones and smart speakers become increasingly ubiquitous, there is growing interest in building automatic speech recognition (ASR) systems that can run directly on-device; end-to-end (E2E) speech recognition models such as recurrent neural network transducers and their variants have recently emerged as prime candidates for this task. Apart from being accurate and compact, such systems need to decode speech with low user-perceived latency (UPL), producing words as soon as they are spoken.
This work examines the impact of various techniques -- model architectures, training criteria, decoding hyperparameters, and endpointer parameters -- on UPL. 
Our analyses suggest that measures of model size (parameters, input chunk sizes), or measures of computation (e.g., FLOPS, RTF) that reflect the model's ability to process input frames are not always strongly correlated with observed UPL.
Thus, conventional algorithmic latency measurements might be inadequate in accurately capturing latency observed when models are deployed on embedded devices.
Instead, we find that factors affecting token emission latency, and endpointing behavior have a larger impact on UPL.
We achieve the best trade-off between latency and word error rate when performing ASR jointly with endpointing, while utilizing the recently proposed alignment regularization mechanism.

\end{abstract}
\noindent\textbf{Index Terms}: speech recognition, latency, on-device, end-to-end, streaming

\section{Introduction}\label{sec:intro}

Automatic speech recognition (ASR) technologies are becoming increasingly ubiquitous in today's technological products; from smartphones to digital assistants on smart speakers, speech has emerged as a prominent input modality~\cite{SchalkwykBeefermanBeaufaysEtAl10, Sarikaya17}.
Some ASR applications (e.g., video captioning, spoken term detection, etc.) admit the use of \emph{offline} recognition where speech can be processed at a later time, through multi-pass systems with bi-directional processing.
However, an important use case -- particularly in the context of digital voice-command assistants, and the focus of this paper -- is the development of ASR systems that can process speech in a \emph{streaming} fashion by producing output hypotheses as speech is input to the system.
Since ASR systems are typically the first stage of a more complex language understanding pipeline~\cite{DeMoriBechetHakkani-TurEtAl08}, it crucial that streaming ASR systems operate with low latency -- i.e., producing words as soon as possible after they are spoken -- in order to ensure that the overall system is responsive to the user's requests.

The demand for low-latency user experiences has driven the speech community to investigate various algorithmic techniques and model architectures.
Recently, end-to-end ASR systems such as connectionist temporal classification (CTC)~\cite{GravesFernandezGomezEtAl06}, and sequence transducers~\cite{graves2012sequence, RaoSakPrabhavalkar17} have emerged as prime candidates for this task since they are compact, accurate, and can be deployed efficiently on embedded devices~\cite{HeSainathPrabhavalkarEtAl19,shangguan2019optimizing,li2019improving}.
The responsiveness of such systems can be measured in terms of two metrics (see Section~\ref{sec:latencydef} for formal definitions): \emph{first token emission delay} -- the time between the start of the user's speech and the first output produced by the system; and \emph{user-perceived latency} (UPL) -- the time between the end of the user's speech and when the ASR system finalizes its hypothesis~\cite{sainath2020streaming}.
Improving these metrics is a multi-faceted problem, and is influenced by all layers of the speech recognition stack: e.g., front-end processing to produce speech frames; model architectures; the endpointer used for end-of-speech detection to close the microphone~\cite{shannon2017improved,chang2019joint}; decoder hyperparameters (e.g., beam size), etc.

Although improving latency is an important research goal, the complexities involved in its measurement have resulted in various definitions and measurement techniques in previous work, which are hard to compare across systems.
Many previous work focus primarily on \emph{algorithmic latency} (e.g.,~ \cite{inaguma2020minimum, li2020high, shi2020emformer, wang2020low} \emph{inter alia}) -- the minimum theoretical amount of time required to process the incoming audio chunk before producing an output (i.e., the time span that the audio chunk represents, along with any required look-ahead frames).
However, such definitions underestimate the true UPL, since they do not account for interactions with other system components such as the endpointer and the decoder.
For the same reason, systems that solely measure latency in terms of the number of floating point operations (FLOPS) in a particular component of the model (e.g., the encoder in~\cite{macoskeybifocal}), do not capture the observed UPL accurately.
Many research work measure endpointer (EP) latency~\cite{mahadeokar2020alignment, yu2020fastemit, sainath2020streaming} -- the difference between when the user stops speaking and when a decision is made to close the microphone, which correlates well with our notion of UPL.
For systems which utilize the ASR result as a part of the endpointing process, or which decide to close the microphone as part of the ASR process~\cite{chang2019joint}, EP latency is directly influenced by the underlying ASR model.
As a final note, some previous work provide a more nuanced view of the latency by measuring it at the sub-utterance level: e.g., model emission delays at the level of frames~\cite{wang2020low}, output tokens~\cite{mahadeokar2020alignment}, words~\cite{wang2020low}, or phrases (segments)~\cite{shangguan2020analyzing,yu2020fastemit} by comparing the models' emission times against ground truth locations in the audio (typically obtained using forced alignments).

In this work, we provide a detailed analysis of the impacts of ASR modeling techniques on UPL by considering two state-of-the-art E2E ASR systems -- the recurrent neural network transducer (RNN-T)~\cite{graves2012sequence,graves2013speech} with LSTM~\cite{hochreiter1997long} encoders; and the recently proposed efficient memory transformer (Emformer)~\cite{shi2020emformer}.
Unlike many previous work, we measure latency in terms of \emph{the actual runtimes on embedded devices under controlled conditions} in order to glean insights that translate directly into measurable improvements in practical applications. Our analyses suggest that measures of model size (parameters, input chunk sizes), or measures of computations (e.g.,FLOPS, real time factor (RTF)) that reflect the model’s ability to process input frames are not always  strongly correlated with observed UPL. Thus, conventional algorithmic latency measurements might be inadequate in accurately capturing latency observed when models are deployed on embedded devices. Instead, we find that factors affecting token emission latency, and endpointing behavior have a much greater impact on UPL. 
We show that modifications to the training loss that improves the speed of the ASR model's token emission result in faster UPL. 
We then consider the impact of different kinds of endpointer strategies on UPL by comparing endpointing based on a fixed amount of acoustic silence~\cite{mahadeokar2020alignment}; a neural endpointer~\cite{shannon2017improved}; and, an integrated E2E endpointer that predicts an end of utterance token as part of the ASR process~\cite{chang2019joint,li2020towards}. Taken together, our results indicate that UPL is dominated by the interactions of two major impact factors: the end-pointer design, and the model's loss function design that promotes faster token emission.
\section{Measuring Latency Metrics}\label{sec:latencydef}

\begin{figure*} 
  \vspace{-2mm} 
 \centering
  \includegraphics[width=0.78\textwidth, height=2.7cm]{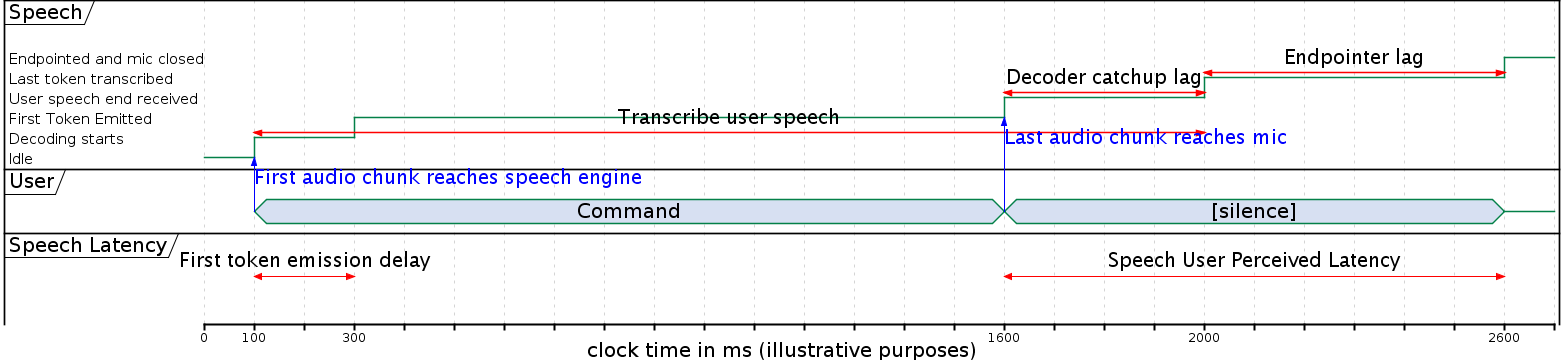}
  \vspace{-2mm}
  \caption{Illustration of timeline and latency metrics of a streaming on-device ASR system.}\label{fig:latencyDEF}
  \vspace{-5mm}
\end{figure*}
 
In this work, we compare ASR systems in terms of their actual runtime behavior when running on an embedded device.
As can be seen in Figure~\ref{fig:latencyDEF}, we measure two speech latency metrics: \emph{first token emission delay} and \emph{user perceived latency}.
The first of these -- \emph{first token emission delay} (200ms, in this example) -- is measured as the time between when the user actually begins speaking (i.e., ignoring any initial silence) and the time when the ASR system actually emits an output token (i.e., non-blank outputs for the transducers in this work). 
The system continues to transcribe incoming speech frames, until a decision is made to close the microphone using an endpointer (either as a standalone system, or jointly trained with the ASR model).
We then measure the second metric -- \emph{user perceived latency} (UPL) -- as the time between when the user finishes speaking and the decision is made to close the microphone (1000ms, in this example).

The UPL is thus a function of both the decoder and the endpointer used to detect the end of the speech. It consists of two factors: \emph{endpointer lag} (600ms, in this example) -- the time required by the endpointer to determine that the user is done speaking; and,~\emph{decoder catchup lag} (400ms, in this example) -- time taken to process any remaining frames received after the user has finished speaking, till the model emits the last non-blank token.
Changing endpointer hyperparameters to operate more aggressively reduces the endpointer lag, at the cost of potentially cutting off users before they are done speaking. Thus, we must tune the endpointer hyperparameters to achieve an acceptable balance between these two desiderata.
The decoder lag, however, is primarily a function of the model's speed to process audio frames and its token emission timing characteristics: e.g., the look-ahead context frames required by the model; the time required to process speech frames relative to the rate at which frames arrive at the device; the intrinsic delay in outputting tokens relative to when they are spoken; etc. Decoder lag can be reduced by changing model architectures to reduce computation, as well as modifying the loss function to encourage the model to output tokens more promptly~\cite{li2020towards}.
Unlike in previous work~\cite{pratap2020scaling, mahadeokar2020alignment} \emph{inter alia}, we examine emission delay of the first token, and the time required to finalize the ASR result without considering per-token emission delays.

\vspace{-2mm}
\subsection{ASR Models}\label{sec:asrmodel}
\vspace{-1mm}
In this work, we study the relationship between ASR accuracy (in terms of word error rate (WER)) and latency metrics in the framework of sequence transducer models~\cite{graves2012sequence, graves2013speech, ZhangLuSakEtAl20}.
In this framework, a model consists of an \emph{encoder} that transforms the input speech frames into a higher-level representation; a \emph{predictor} that models the sequence of previously predicted output tokens; and a \emph{joiner} that combines the information from the encoder and the predictor to produce a distribution over output tokens. 
As a result of their compact size and ease of deployment, the sequence transducer framework has been used widely for accurate on-device ASR~\cite{HeSainathPrabhavalkarEtAl19, sainath2020streaming, shangguan2020analyzing, shangguan2019optimizing}.
In this work, following~\cite{mahadeokar2020alignment}, we process the input speech into 80-dimensional log Mel-filterbank features, computed using a 25ms window with a 10ms frame shift.
All models output a distribution over 4096 tokens (including the special `blank' token) from a pre-trained sentence-piece model~\cite{kudo2018sentencepiece}.

We conduct our experiments with an recurrent neural network transducer (RNN-T) model that uses unidirectional long short term memory (LSTM) cells~\cite{hochreiter1997long, sak2014long} in the encoder. 
More specifically, the LSTM cells employ a projection layer to restrict weight matrix sizes, along with layer normalization~\cite{ba2016layer} to improve stability during training. 
We also compare the RNN-T model with the recently proposed efficient memory transformer (Emformer)~\cite{shi2020emformer} which uses transformers with an augmented memory that captures information from previous parts of the utterance. The Emformer~\cite{shi2020emformer}, can be configured by specifying the number of attention heads ($|H|$), which compute attention over keys, values, and queries of dimension ($|W|$), and the size of the input audio chunks ($|C|$).
We refer the reader to~\cite{shi2020emformer} for more details on the Emformer.

In both the RNN-T and the Emformer model, we use time-reduction layers~\cite{alvarez2016efficient} between encoder layers to concatenate and project features and cell activation values from the consecutive frames in the previous layer, before feeding them to the next layer.
We refer to this technique as \emph{model striding} in this work; thus, a model with two time-reduction layers of size 2 and 3, would correspond to an overall stride of 6.

\vspace{-2mm}
\subsection{Factors Impacting Latency Metrics}
\vspace{-1mm} 
We categorize the factors contributing to latency below.
Although explained in the context of the sequence transducer framework, these factors also apply to other model frameworks.

\vspace{-2mm}
\subsubsection{Model Design}
\vspace{-1mm} 

The model structure and the training loss function impact token emission delays, thus decoder catchup lag and UPL.
\vspace{-1mm}

\begin{enumerate}[leftmargin=0cm,itemindent=.5cm,labelwidth=\itemindent,labelsep=0cm,align=left]
    \item \textbf{Cell architecture} is the building block of each model layer.
    Besides LSTMs, Emformers and their variants, common cell architectures in ASR also include convolutional neural networks (CNNs)~\cite{abdel2014convolutional, gulati2020conformer} and transformers~\cite{chan2015listen,wu2020streaming}.
    
    \vspace{-1mm} 
    \item \textbf{Model layer configuration} is the design of models in terms of layer width, number of layers.
    On-device ASR models are typically evaluated after quantization~\cite{shangguan2019optimizing}, as is the case in this work, which improves latency; factors such as sparsity~\cite{shangguan2019optimizing}, however, are not explored in this work.

    \vspace{-1mm} 
    \item \textbf{Model stride and chunk size} determine the number of contextual audio frames needed for the model to compute one forward pass.
    Some models need past and look-ahead contextual audio information during their forward passes.
    In Emformers, for example, the amount of memory carried over from the previous layer and the number of audio frames per step constitute the \textit{chunk size $|C|$} of the model.
    Model strides (defined in Section~\ref{sec:asrmodel}) also have a large impact: larger model strides and chunk sizes lead to longer algorithmic latency because the ASR model needs to wait to accumulate enough audio frames at each step of the computation.
    
    \vspace{-1mm} 
    \item \textbf{Loss function design} impacts the UPL if it is modeled as part of the loss function. Some work rely on token-alignment, and apply early- or late-penalties to token emissions that differ from ground truth emission times~\cite{mahadeokar2020alignment,li2020towards}; others modify the model training loss to encourage faster token emission~\cite{yu2020fastemit}.
    In this paper, we compare FastEmit RNN-T loss~\cite{yu2020fastemit} with Alignment Restricted RNN-T loss (ArRNN-T)~\cite{mahadeokar2020alignment}, and the original RNN-T loss~\cite{graves2012sequence}, to determine the impact of these losses on latency metrics.
\end{enumerate}

\vspace{-4mm}
\subsubsection{Endpointer}
\vspace{-1mm}
The endpointer model runs in parallel with the decoder and detects the end-of-query in order to close the microphone.
In our paper, we benchmark three endpointers. 
\vspace{-1mm}

\begin{enumerate}[leftmargin=0cm,itemindent=.5cm,labelwidth=\itemindent,labelsep=0cm,align=left]   
    \item \textbf{Static endpointer} (StaticEP) determines the end-of-query by tracking the trailing silence from the last emitted non-blank token of the ASR model.
    We denote \textit{staticEP N(s)} as a static endpointer that closes the mic after \textit{N(s)} of trailing silence.
    \vspace{-1mm}

    \item \textbf{Neural endpointer} (NEP) takes in acoustic features and predicts end-of-speech using a light-weight LSTM model \cite{shannon2017improved}.
    This neural endpointer is trained to predict binary outputs at each frame.
    We smooth the output using a small sliding window to improve robustness.
    The smoothed output is then compared against a threshold to make an endpoint decision -- the lower the threshold, the more aggressive the neural endpointer.
    \vspace{-1mm}

    \item \textbf{E2E endpointer} (E2E-EP) utilizes the ASR model to predict end-of-speech ($\langle\text{eos}\rangle$) label directly~\cite{chang2019joint,li2020towards,yu2020fastemit}.
    In the RNN-T training phase, an end-of-speech label is added to the end of each utterance's reference; during decoding, an endpoint is detected if the $\langle\text{eos}\rangle$ label is predicted in the best hypothesis with confidence exceeding a certain threshold.
\end{enumerate}

\vspace{-4mm}
\subsection{Other Factors}
\vspace{-1mm}
Inference in the neural transducer framework is performed using an approximate beam search~\cite{graves2012sequence, jain2020rnnt}.
Parameters such as \textbf{beam size} (i.e., the maximum number of hypotheses, i.e. model states, that are expanded at every step of decoding), affect the overall computation and thus, UPL.
However, in practice, most of the model computation is dominated by processing in the encoder, and the decoder beam size has a relatively small impact on UPL for reasonable beam sizes (e.g., $\leq$15).  
Some ASR systems include \textbf{additional models}, e.g., second pass rescoring that rescores the first-pass ASR outputs~\cite{li2020towards,sainath2020streaming}, or on-the-fly biasing with language models~\cite{Le2021deepshallow,kim2020improved,HeSainathPrabhavalkarEtAl19} that improves WER via fusion.
Furthermore, ASR models often run on a multi-core, multi-threaded \textbf{computational environment}, possibly with dedicated neural network accelerators such as GPUs or TPUs to improve latency.
Latency changes due to second-pass models and the computational environment are outside of the scope of this paper; we focus on first-pass, single-core CPU-based model evaluation on Android devices.

\vspace{-1mm}
\section{Experiments}
\subsection{Voice Command Data}
\vspace{-1mm}
We analyze the factors that affect the UPL on models trained and evaluated with an in-house Voice Command dataset.
Recognizing voice commands is one of the most popular and latency-demanding use cases of on-device ASR. Users expect fast responses from the ASR models when they ask to turn on the lights.
We use the same voice command dataset described in~\cite{mahadeokar2020alignment}.
It comes from two sources: a 12.5k-hour human-transcribed speech collected via mobile devices by 20k crowd-sourced workers; and, a 1k-hour smart speaker voice commands collection from the production traffic.
In both of the data sources, personally identifiable information (PII) is removed, and the sound is morphed, speed perturbed~\cite{ko2015audio} at 0.9x and 1.1x the original speech.
We further modify the data with simulated reverberation and background noise, which is collected from publicly available videos.

Our WER evaluation data consists of 10k hand-transcribed utterances collected from volunteer participants in a Smart Speaker’s in-house pilot program.
These utterances are also anonymized, with PII removed.
We randomly select a subset of 100 utterances from the evaluation dataset which are used to compute the latency benchmarks described in Section~\ref{sec:latencydef} on high-end Android devices; metrics are reported as the averages of two independent runs.
These benchmark utterances contain human-transcribed endpoints, indicating the audio time stamp of the last token.
We append 2 seconds of silence to the end of all utterances, to more accurately measure ASR endpointing behavior.

\vspace{-2mm}
\subsection{Results and Discussions}\vspace{-1mm}
We first examine the impact of algorithmic model design on the UPL.
Table~\ref{tab:latencyModel} shows results of 4 models trained with RNN-T loss, and evaluated with the same endpointer set up -- a StaticEP with 0.9s silence allowed after the last emitted token.
Both Emformer- and LSTM-based RNN-Ts here are stride 8 models.
The predictor of the RNN-Ts are built with LSTM cells.
We calculate the FLOPS of the models assuming $|$audio frames$|/|$wpm tokens$|$=8.
The model sizes are reported after int8 quantization: $1\text{MB}\approx1\text{M}$ parameters from the model.
\begin{table}[t]
  \caption{Comparing UPL and model architecture.
  The predictors are constructed with LSTM cells.
  Encoder and predictor layers are denoted as (number of layers) x (number of cells per layer). P50: median percentile.}
    \vspace{-2mm}
  \label{tab:latencyModel}
  \centering
  \begin{tabular}{|@{\hspace{2pt}}p{0.16cm}|@{\hspace{2pt}}p{1.78cm}|@{\hspace{2pt}}p{0.7cm}|@{\hspace{2pt}}p{0.55cm}|@{\hspace{2pt}}p{0.8cm}|@{\hspace{2pt}}p{0.58cm}|@{\hspace{2pt}}p{0.75cm}|@{\hspace{2pt}}p{0.4cm}|}
  \toprule
    \multicolumn{8}{c}{LSTM-based RNN-T Encoder Cells}   \\\hline
    ID & Enc           &   Pred     & size        & P50  & WER    & FLOPS & RTF\\ 
       & layers        &  layers    & (MB)        & UPL(s) &      & mil& P50\\\hline
    A & 8x640          &   2x512   &  69.9        & 1.60    & 1.94    & 33.8& 0.53 \\\hline
    B & 8x320+4x640    &   2x512    &  57.1       & 1.47   & 2.25    & 26.3& 0.42\\\hline
    C & 8x640          &   4x512    &  91.4       & 1.57   & 2.06    &36.5 & 0.55\\\hline
    \multicolumn{8}{c}{Emformer-based RNN-T Encoder Cells}   \\\hline
    D & \footnotesize{20 x (8$|H|$, 512$|W|$, 5$|C|$)}&   2x512   & 77.4 &  1.50&  2.14  &  11.8& 0.21  \\\hline
  \end{tabular}
  \vspace{-6mm}
\end{table}

As can be seen in Table~\ref{tab:latencyModel}, conventional measures of the model's theoretical speed in processing input audio chunks do not correlate well with observed UPL, especially when we compare RNN-T constructed with different cells. Model C, which is 31\% larger than Model A, has a similar median UPL.
Comparing models in terms of FLOPS or RTF (real-time factor: the ratio of the time required to process audio frames relative to the length of the audio), we observe that a 55\% reduction in FLOPS, or a 50\% RTF improvement, from Model B to Model D actually results in 2\% \emph{degradation} in median UPL. Similarly, in terms of model stride and chunk sizes, which characterize the model's algorithmic latency, a 100\% increase in model stride only results in 5\% improvement in the UPL of the model comparing Model F to Model G (Figure~\ref{fig:latencyBig}, Table~\ref{tab:latencyfinaltable}); Model A and B have the same stride and chunk sizes, and thus, the same algorithmic latency, but 8\% different UPL. Note that all models in Table~\ref{tab:latencyModel} have RTF$<$1. For models with RTF$\geq$1, the decoder catchup lag becomes significant, thus making UPL more sensitive to RTF and model chunk/stride size changes.

If algorithmic latency is not a good proxy for the model's UPL, what is a good measure then?
The model's speed in emitting tokens has a strong impact on the UPL.
To illustrate this, we take model A from Table~\ref{tab:latencyModel} and train it with three different loss functions. We swept $\lambda$ for the FastEmit RNN-T loss setting, and chose 0.004~\cite{yu2020fastemit}; we swept the right buffer size of the alignment-restricted RNN-T loss setting and chose $\text{r-buffer}$= 6~\cite{mahadeokar2020alignment}.
We present results after evaluating models with a StaticEP configured to trigger after 0.9s of trailing silence (i.e., blank outputs) in Table~\ref{tab:latencyloss}. Note that the EP lag in the table measures the time from a ground-truth human annotated end-of-speech to the time that the ASR closes the mic. As can be seen in the table, the alignment restriction in the ArRNN-T loss function reduces $1^{st}$ token delay by $19\%$ from 0.62s to 0.50s.
As a result of the earlier token emission, the endpointer lag of the staticEP also improves, which improves median UPL $14\%$.
This UPL improvement with ArRNN-T loss, however, comes at the cost of $14\%$ degradation of WER.
FastEmit, on the other hand, did not seem to improve UPL in this particular experiment.

\begin{table}[h]
\caption{Comparing Model A trained with RNN-T, ArRNN-T and FastEmit losses. P(N): the $N^{th}$ percentile; ``Dcdr Ctchp": decoder catchup lag(s); ``Tk delay": 1$^{st}$ token delay(s). }  
\vspace{-3mm}
\label{tab:latencyloss}
\centering
  \begin{tabular}{|@{\hspace{2pt}}p{0.16cm}|@{\hspace{1.6pt}}p{0.69cm}|@{\hspace{1.4pt}}p{1.0cm}|@{\hspace{1.45pt}}p{0.25cm}|@{\hspace{1.45pt}}p{0.25cm}|@{\hspace{1.45pt}}p{0.25cm}|@{\hspace{1.45pt}}p{0.25cm}|@{\hspace{1.45pt}}p{0.54cm}|@{\hspace{1.8pt}}p{0.56cm}|}
  \toprule
    ID& Tk delay & \multicolumn{1}{l}{Loss} & \multicolumn{2}{c}{UPL(s)}  &  \multicolumn{2}{c}{Dcdr Ctchp} & EP lag & WER \\
   \cmidrule(lr){4-5}\cmidrule(lr){6-7}
    & P50 &   & \multicolumn{1}{l}{P50}& \multicolumn{1}{l}{P90} & \multicolumn{1}{l}{P50} &\multicolumn{1}{l}{P90}   & P50 &  \\\hline
   A & 0.62 & RNNT     & 1.60 & 1.88  & 0.56  & 0.75   & 1.00 & 1.94 \\\hline
   E & 0.56 & FastEmit & 1.60 & 1.85  & 0.56  & 0.75   & 1.03 & 2.04 \\\hline
   J1& 0.50 & ArRNNT   & 1.37 & 1.62  & 0.62  & 0.78   & 0.81 & 2.22 \\\hline
  \end{tabular}
    \vspace{-2mm}
\end{table}

\begin{figure}[h]
  \includegraphics[width=0.4\textwidth]{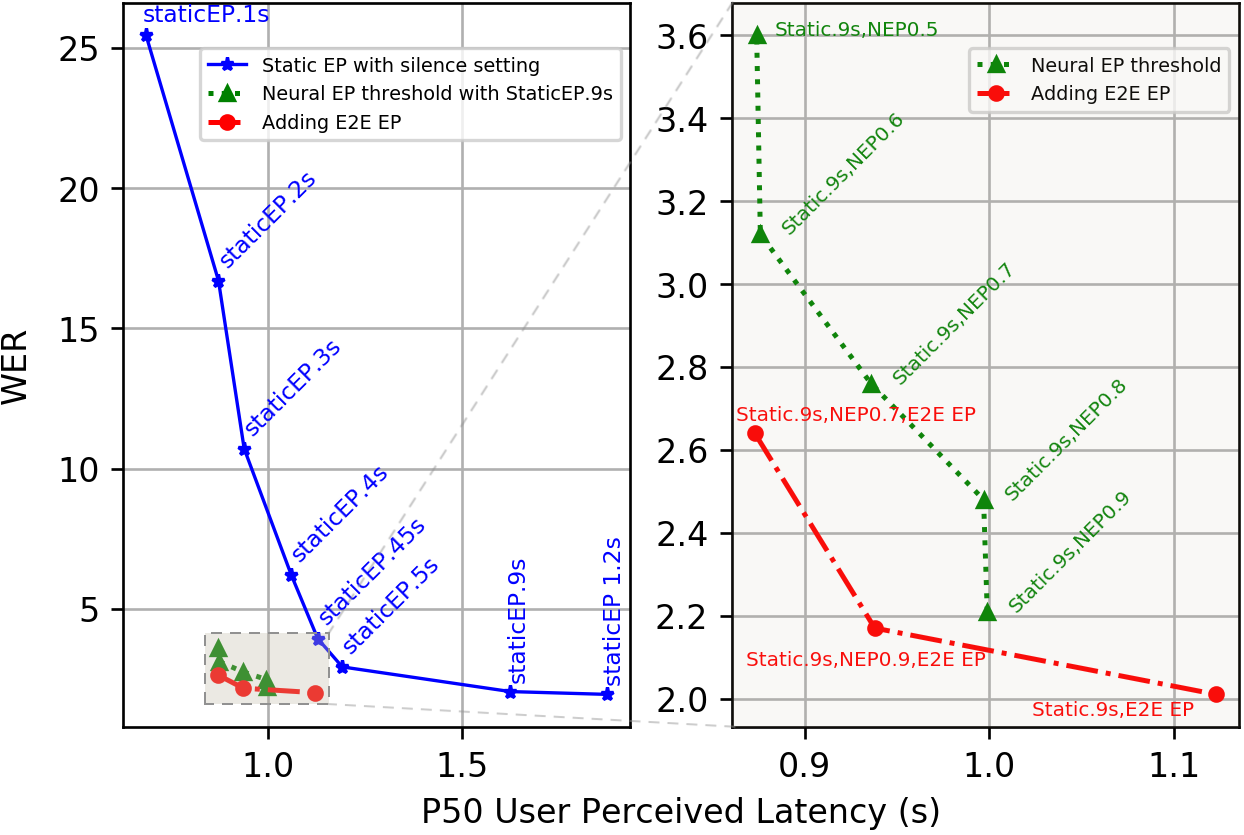}
  \vspace{-2mm}
  \caption{WER vs. 50$^{th}$ percentile UPL for Model A. Left: EP is StaticEP with various silence settings. Right: zoomed-in version of StaticEP.9s with additional NEP and E2E-EP.}\label{fig:latencyEP}
  \vspace{-1mm}
\end{figure}

The most significant component of the UPL is the endpointer lag. One can observe from Table~\ref{tab:latencyloss} that EP lag is almost $\geq50\%$ of the overall UPL in a RNN-T model. By adopting a different endpointing model, we can further optimize the WER-latency trade-offs of Model A, shown in Figure~\ref{fig:latencyEP}. The left sub-plot shows the different StaticEP settings for the amount of silence tolerated after the final token; the right sub-plot shows how neural endointer (NEP) and end-to-end endpointer (E2E-EP) impact model WER and latency in a more close up look. 

\begin{table}[ht]
\vspace{-1mm}
\begin{minipage}[h]{1.0\linewidth}
  \centering
    \captionof{table}{Model striding, loss function and the EP impact RNN-T models. \textbf{S}: StaticEP0.9s; \textbf{N}: NEP 0.7 threshold; \textbf{E}: E2E-EP modeled by ASR; ``Dcdr Ctchp": P50 decoder catchup lag(s).}\vspace{-2mm}
  \begin{tabular}{|@{\hspace{2pt}}p{0.16cm}|@{\hspace{2pt}}p{0.55cm}|@{\hspace{2pt}}p{0.45cm}|@{\hspace{2pt}}p{0.6cm}|@{\hspace{2pt}}p{0.6cm}||@{\hspace{2pt}}p{0.16cm}|@{\hspace{2pt}}p{0.55cm}|@{\hspace{2pt}}p{0.45cm}|@{\hspace{2pt}}p{0.6cm}|@{\hspace{2pt}}p{0.25cm}|}\hline 
  \toprule
    \multicolumn{5}{c}{RNN-T with StaticEP0.9s}   &  \multicolumn{5}{c}{ArRNN-T Stride8 w/ EPs}   \\\hline
    ID &Stride & UPL  & Dcdr & $EP_\text{lag}$  & ID  & EP  & UPL  & Dcdr & $EP_\text{lag}$ \\ 
       &      & P50      & Ctchp & P50 &     &     & P50  & Ctchp& P50 \\\hline 
    F  &  2   &  1.81    & 0.84  & 0.94 & J1  & \textbf{S}   & 1.37 & 0.62  & 0.81   \\\hline
    G  &  4   &  1.72    & 0.69  & 1.06 &  J2  & \textbf{SN}  & 0.94 & 0.63  & 0.19 \\\hline
    H  &  6   &  1.60    & 0.63  & 0.97 &  K   & \textbf{SNE} & 0.87 & 0.62  & 0.13   \\\hline
    A  &  8   &  1.60    & 0.56  & 1.00 &  \multicolumn{5}{c|}{} \\\hline
    I  &  12  &  1.35    & 0.53  & 0.79  &  \multicolumn{5}{c|}{}   \\\hline
  \end{tabular}
    \label{tab:latencyfinaltable}
\end{minipage}\hfill
\vspace{2mm}
\begin{minipage}[h]{1\linewidth}
\centering
 \includegraphics[width=0.8\linewidth]{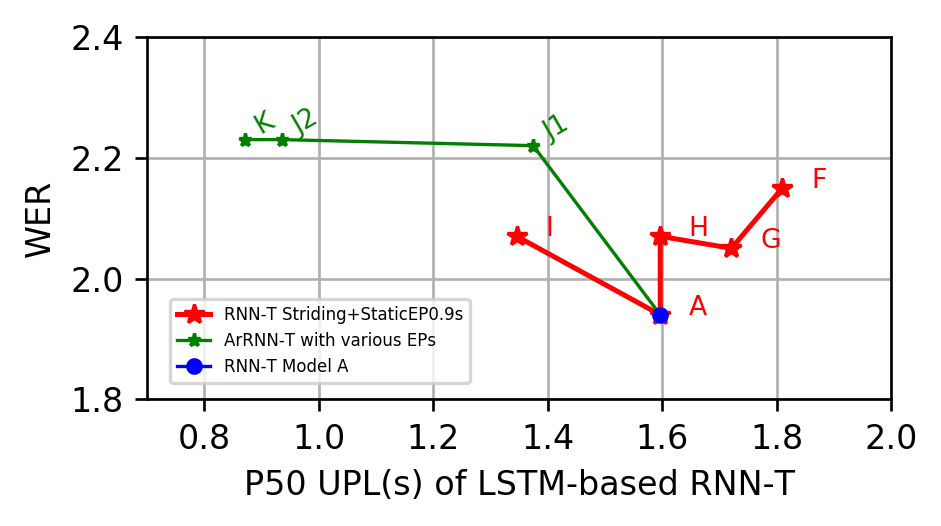}
 \captionof{figure}{WER vs. P50 UPL curves influenced by model striding, loss function and EP design in Table~\ref{tab:latencyfinaltable}.}\label{fig:latencyBig}
\label{fig:image}
\vspace{-5mm}
\end{minipage}
\end{table} 

As shown in Figure~\ref{fig:latencyEP}, the shorter the amount of silence allowed after the last emitted non-blank token, the larger the WER degradation caused by an increase in deletion errors due to premature end-pointing. 
Using more accurate endpointers, such as NEP and E2E-EP, the ASR model achieves a much better trade-off curve between WER and latency. 

Since endpointer design effectively reduces endpointer lags, model striding design reduces decoder catchup lags, and the model loss function design impacts both, we show the impact of these factors on the UPL in the same graph (Figure~\ref{fig:latencyBig} and Table~\ref{tab:latencyfinaltable}) to illustrate their relative importance. 
In our experiments, we find that the best WER-latency trade-off is achieved by modeling ArRNN-T loss together with endpointing, using an E2E endpointer (Model K).

\vspace{-1mm}
\section{Conclusions}
\vspace{-1mm}
In this paper, we study factors which impact user-perceived latency in speech which consists of decoder catchup and endpointer lags.
We analyze the impact of ASR model design choices on the UPL: endpointer models, model algorithmic designs, decoder parameters and other pipeline artifacts. We show that, counter-intuitively, metrics that correlate with how fast the ASR model process audio frames: model sizes, RTF, model computations (FLOPS) do not always strongly correlate with UPL. Instead, we use controlled experiments to show that factors affecting token emission latency, and endpointing behavior have much greater impact on UPL. We achieve the best trade-off between latency and word error rate when performing ASR jointly with endpointing, and using the recently proposed alignment restriction loss.

\bibliographystyle{IEEEtran}

\bibliography{mybib}

\begin{thebibliography}{10}
\providecommand{\url}[1]{#1}
\csname url@samestyle\endcsname
\providecommand{\newblock}{\relax}
\providecommand{\bibinfo}[2]{#2}
\providecommand{\BIBentrySTDinterwordspacing}{\spaceskip=0pt\relax}
\providecommand{\BIBentryALTinterwordstretchfactor}{4}
\providecommand{\BIBentryALTinterwordspacing}{\spaceskip=\fontdimen2\font plus
\BIBentryALTinterwordstretchfactor\fontdimen3\font minus
  \fontdimen4\font\relax}
\providecommand{\BIBforeignlanguage}[2]{{%
\expandafter\ifx\csname l@#1\endcsname\relax
\typeout{** WARNING: IEEEtran.bst: No hyphenation pattern has been}%
\typeout{** loaded for the language `#1'. Using the pattern for}%
\typeout{** the default language instead.}%
\else
\language=\csname l@#1\endcsname
\fi
#2}}
\providecommand{\BIBdecl}{\relax}
\BIBdecl

\bibitem{SchalkwykBeefermanBeaufaysEtAl10}
J.~Schalkwyk, D.~Beeferman, F.~Beaufays, B.~Byrne, C.~Chelba, M.~Cohen,
  M.~Kamvar, and B.~Strope, ``{``Your word is my command”: Google search by
  voice: A case study},'' in \emph{Advances in speech recognition}, 2010, pp.
  61--90.

\bibitem{Sarikaya17}
R.~{Sarikaya}, ``The technology behind personal digital assistants: An overview
  of the system architecture and key components,'' \emph{IEEE Signal Processing
  Magazine}, vol.~34, no.~1, pp. 67--81, 2017.

\bibitem{DeMoriBechetHakkani-TurEtAl08}
R.~{De Mori}, F.~{Bechet}, D.~{Hakkani-Tur}, M.~{McTear}, G.~{Riccardi}, and
  G.~{Tur}, ``Spoken language understanding,'' \emph{IEEE Signal Processing
  Magazine}, vol.~25, no.~3, pp. 50--58, 2008.

\bibitem{GravesFernandezGomezEtAl06}
A.~Graves, S.~Fern{\'a}ndez, F.~Gomez, and J.~Schmidhuber, ``{Connectionist
  Temporal Classification: Labelling Unsegmented Sequence Data with Recurrent
  Neural Networks},'' in \emph{Proc. of ICML}, 2006.

\bibitem{graves2012sequence}
A.~Graves, ``Sequence transduction with recurrent neural networks,'' in
  \emph{ICML Representation Learning Workshop}, 2012.

\bibitem{RaoSakPrabhavalkar17}
K.~{Rao}, H.~{Sak}, and R.~{Prabhavalkar}, ``Exploring architectures, data and
  units for streaming end-to-end speech recognition with rnn-transducer,'' in
  \emph{Proc. of ASRU}, 2017.

\bibitem{HeSainathPrabhavalkarEtAl19}
Y.~{He}, T.~N. {Sainath}, R.~{Prabhavalkar}, I.~{McGraw}, R.~{Alvarez},
  D.~{Zhao}, D.~{Rybach}, A.~{Kannan}, Y.~{Wu}, R.~{Pang}, Q.~{Liang},
  D.~{Bhatia}, Y.~{Shangguan}, B.~{Li}, G.~{Pundak}, K.~C. {Sim}, T.~{Bagby},
  S.~{Chang}, K.~{Rao}, and A.~{Gruenstein}, ``Streaming end-to-end speech
  recognition for mobile devices,'' in \emph{Proc. of {ICASSP}}, 2019.

\bibitem{shangguan2019optimizing}
Y.~Shangguan, J.~Li, Q.~Liang, R.~Alvarez, and I.~McGraw, ``Optimizing speech
  recognition for the edge,'' in \emph{MLSys On-device Intelligence Workshop},
  2020.

\bibitem{li2019improving}
J.~Li, R.~Zhao, H.~Hu, and Y.~Gong, ``Improving {RNN} transducer modeling for
  end-to-end speech recognition,'' in \emph{Proc. of ASRU}, 2019.

\bibitem{sainath2020streaming}
T.~N. Sainath, Y.~He, B.~Li, A.~Narayanan, R.~Pang, A.~Bruguier, S.-Y. Chang,
  W.~Li, R.~Alvarez, Z.~Chen \emph{et~al.}, ``A streaming on-device end-to-end
  model surpassing server-side conventional model quality and latency,'' in
  \emph{Proc. of ICASSP}, 2020.

\bibitem{shannon2017improved}
M.~Shannon, G.~Simko, S.-Y. Chang, and C.~Parada, ``Improved end-of-query
  detection for streaming speech recognition.'' in \emph{Proc. of Interspeech},
  2017.

\bibitem{chang2019joint}
S.-Y. Chang, R.~Prabhavalkar, Y.~He, T.~N. Sainath, and G.~Simko, ``Joint
  endpointing and decoding with end-to-end models,'' in \emph{Proc. of ICASSP},
  2019.

\bibitem{inaguma2020minimum}
H.~Inaguma, Y.~Gaur, L.~Lu, J.~Li, and Y.~Gong, ``Minimum latency training
  strategies for streaming sequence-to-sequence asr,'' in \emph{Proc. of
  ICASSP}, 2020.

\bibitem{li2020high}
J.~Li, R.~Zhao, E.~Sun, J.~H. Wong, A.~Das, Z.~Meng, and Y.~Gong,
  ``High-accuracy and low-latency speech recognition with two-head contextual
  layer trajectory lstm model,'' in \emph{Proc. of ICASSP}, 2020.

\bibitem{shi2020emformer}
Y.~Shi, Y.~Wang, C.~Wu, C.~Yeh, J.~Chan, F.~Zhang, D.~Le, and M.~L. Seltzer,
  ``{Emformer: Efficient Memory Transformer Based Acoustic Model For Low
  Latency Streaming Speech Recognition},'' in \emph{Proc. of ICASSP}, 2021.

\bibitem{wang2020low}
C.~Wang, Y.~Wu, L.~Lu, S.~Liu, J.~Li, G.~Ye, and M.~Zhou, ``{Low Latency
  End-to-End Streaming Speech Recognition with a Scout Network},'' in
  \emph{Proc. of Interspeech}, 2020.

\bibitem{macoskeybifocal}
J.~Macoskey, G.~P. Strimel, and A.~Rastrow, ``Bifocal neural {ASR}: Exploiting
  keyword spotting for inference optimization,'' in \emph{Proc. of ICASSP},
  2021.

\bibitem{mahadeokar2020alignment}
J.~{Mahadeokar}, Y.~{Shangguan}, D.~{Le}, G.~{Keren}, H.~{Su}, T.~{Le}, C.~F.
  {Yeh}, C.~{Fuegen}, and M.~L. {Seltzer}, ``Alignment restricted streaming
  recurrent neural network transducer,'' in \emph{IEEE Spoken Language
  Technology Workshop}, 2021.

\bibitem{yu2020fastemit}
J.~Yu, C.-C. Chiu, B.~Li, S.-y. Chang, T.~N. Sainath, Y.~He, A.~Narayanan,
  W.~Han, A.~Gulati, Y.~Wu \emph{et~al.}, ``Fastemit: Low-latency streaming asr
  with sequence-level emission regularization,'' in \emph{Proc. of ICASSP},
  2021.

\bibitem{shangguan2020analyzing}
Y.~Shangguan, K.~Knister, Y.~He, I.~McGraw, and F.~Beaufays, ``{Analyzing the
  Quality and Stability of a Streaming End-to-End On-Device Speech
  Recognizer},'' in \emph{Proc. of Interspeech}, 2020.

\bibitem{graves2013speech}
A.~Graves, A.-R. Mohamed, and G.~Hinton, ``Speech recognition with deep
  recurrent neural networks,'' in \emph{Proc. of ICASSP}, 2013.

\bibitem{hochreiter1997long}
S.~Hochreiter and J.~Schmidhuber, ``Long short-term memory,'' \emph{Neural
  computation}, vol.~9, no.~8, pp. 1735--1780, 1997.

\bibitem{li2020towards}
B.~Li, S.-Y. Chang, T.~N. Sainath, R.~Pang, Y.~He, T.~Strohman, and Y.~Wu,
  ``Towards fast and accurate streaming end-to-end {ASR},'' in \emph{Proc. of
  ICASSP}, 2020.

\bibitem{pratap2020scaling}
V.~Pratap, Q.~Xu, J.~Kahn, G.~Avidov, T.~Likhomanenko, A.~Hannun,
  V.~Liptchinsky, G.~Synnaeve, and R.~Collobert, ``Scaling up online speech
  recognition using {ConvNets},'' in \emph{Proc. of Interspeech}, 2020.

\bibitem{ZhangLuSakEtAl20}
Q.~{Zhang}, H.~{Lu}, H.~{Sak}, A.~{Tripathi}, E.~{McDermott}, S.~{Koo}, and
  S.~{Kumar}, ``{Transformer Transducer: A Streamable Speech Recognition Model
  with Transformer Encoders and RNN-T Loss},'' in \emph{Proc. of ICASSP}, 2020.

\bibitem{kudo2018sentencepiece}
T.~Kudo and J.~Richardson, ``{S}entence{P}iece: A simple and language
  independent subword tokenizer and detokenizer for neural text processing,''
  in \emph{Proc. of EMNLP}, 2018.

\bibitem{sak2014long}
H.~Sak, A.~W. Senior, and F.~Beaufays, ``Long short-term memory recurrent
  neural network architectures for large scale acoustic modeling,'' in
  \emph{Proc. of Interspeech}, 2014.

\bibitem{ba2016layer}
J.~L. Ba, J.~R. Kiros, and G.~E. Hinton, ``Layer normalization,'' in
  \emph{Proc. of NeurIPS}, 2016.

\bibitem{alvarez2016efficient}
R.~Alvarez, R.~Prabhavalkar, and A.~Bakhtin, ``On the efficient representation
  and execution of deep acoustic models,'' in \emph{Proc. of Interspeech},
  2016.

\bibitem{abdel2014convolutional}
O.~Abdel-Hamid, A.-r. Mohamed, H.~Jiang, L.~Deng, G.~Penn, and D.~Yu,
  ``Convolutional neural networks for speech recognition,'' \emph{IEEE/ACM
  Transactions on audio, speech, and language processing}, vol.~22, no.~10, pp.
  1533--1545, 2014.

\bibitem{gulati2020conformer}
A.~Gulati, J.~Qin, C.-C. Chiu, N.~Parmar, Y.~Zhang, J.~Yu, W.~Han, S.~Wang,
  Z.~Zhang, Y.~Wu \emph{et~al.}, ``Conformer: Convolution-augmented transformer
  for speech recognition,'' in \emph{Proc. of Interspeech}, 2020.

\bibitem{chan2015listen}
W.~{Chan}, N.~{Jaitly}, Q.~{Le}, and O.~{Vinyals}, ``Listen, attend and spell:
  A neural network for large vocabulary conversational speech recognition,'' in
  \emph{Proc. of ICASSP}, 2016.

\bibitem{wu2020streaming}
C.~Wu, Y.~Wang, Y.~Shi, C.-F. Yeh, and F.~Zhang, ``{Streaming Transformer-Based
  Acoustic Models Using Self-Attention with Augmented Memory},'' in \emph{Proc.
  of Interspeech}, 2020.

\bibitem{jain2020rnnt}
M.~Jain, K.~Schubert, J.~Mahadeokar, C.-F. Yeh, K.~Kalgaonkar, A.~Sriram,
  C.~Fuegen, and M.~L. Seltzer, ``{RNN-T} for latency controlled {ASR} with
  improved beam search,'' \emph{arXiv preprint arXiv:1911.01629}, 2019.

\bibitem{Le2021deepshallow}
D.~Le, G.~Keren, J.~Chan, J.~Mahadeokar, C.~Fuegen, and M.~L. Seltzer, ``{Deep
  Shallow Fusion for RNN-T Personalization},'' in \emph{IEEE Spoken Language
  Technology Workshop}, 2021.

\bibitem{kim2020improved}
S.~Kim, Y.~Shangguan, J.~Mahadeokar, A.~Bruguier, C.~Fuegen, M.~L. Seltzer, and
  D.~Le, ``Improved neural language model fusion for streaming recurrent neural
  network transducer,'' in \emph{Proc. of ICASSP}, 2021.

\bibitem{ko2015audio}
T.~Ko, V.~Peddinti, D.~Povey, and S.~Khudanpur, ``Audio augmentation for speech
  recognition,'' in \emph{Proc. of Interspeech}, 2015.

\end{thebibliography}

\end{document}